# Quantifying the risk of workplace COVID-19 clusters in terms of commuter, workplace, and population characteristics

Christopher E. Overton*[1,2,3], Rachel Abbey[4], Tarrion Baird[2,5], Rachel Christie[2,4], Owen Daniel[6], Julie Day[2,4], Matthew Gittins[7], Owen Jones[2], Robert Paton[2], Maria Tang[2], Tom Ward[2], Jack Wilkinson[7], Camilla Woodrow-Hill[2,8], Tim Aldridge[9], Yiqun Chen[9]

*corresponding author: c.overton@liverpool.ac.uk.
[1]Department of Mathematical Sciences, University of Liverpool, Liverpool, L69 7ZL, UK
[2]All Hazards Intelligence, Data, Analytics and Surveillance, UK Health Security Agency, London, SW1P 3JR, UK
[3]Department of Mathematics, University of Manchester, Manchester, M13 9PL, UK
[4]Analytics and Data Science, Data, Analytics and Surveillance, UK Health Security Agency, London, SW1P 3JR, UK
[5]Department of Pathology, University of Cambridge, Cambridge, CB2 1QP, UK
[6]Data Science Campus, Office for National Statistics, UK
[7]Division of Population Health, Health Services Research & Primary Care, University of Manchester, Manchester, M13 9NT, UK
[8]Division of Psychology Communication and Human Neuroscience, University of Manchester, Manchester, M13 9NT, UK
[9]Science Division, Health and Safety Executive, Buxton, SK17 9JN, UK

**Abstract**

*Objectives*
To identify and quantify risk factors that contribute to clusters of COVID-19 in the workplace.

*Methods*
We identified clusters of COVID-19 cases in the workplace and investigated the characteristics of the individuals, the workplaces, the areas they work, and the methods of commute to work, through data linkages based on Middle Layer Super Output Areas (MSOAs) in England between 20/06/2021 and 20/02/2022. We estimated associations between potential risk factors and workplace clusters, adjusting for plausible confounders identified using a Directed Acyclic Graph (DAG).

*Results*
For most industries, increased physical proximity in the workplace was associated with increased risk of COVID-19 clusters, while increased vaccination was associated with reduced risk. Commuter demographic risk factors varied across industry, but for the majority of industries, a higher proportion of black/african/caribbean ethnicities, and living in deprived areas, was associated with increased cluster risk. A higher proportion of commuters in the 60-64 age group was associated with reduced cluster risk. There were significant associations between gender, work commute methods, and staff contract type with cluster risk, but these were highly variable across industries.

Conclusions
This study has used novel national data linkages to identify potential risk factors of workplace COVID-19 clusters, including possible protective effects of vaccination and increased physical distance at work. The same methodological approach can be applied to wider occupational and environmental health research.

## 1. Introduction
1.1. Background/rationale

Throughout the COVID-19 pandemic, outbreaks of cases in workplaces have frequently been reported worldwide [1], [2]. Some industries have seen elevated rates of outbreaks, often including healthcare settings, where the risk of exposure to patients with COVID-19 is high, and essential services, where work is unable to be carried out from home. The impact of workplace outbreaks is two-fold. Firstly, workplace outbreaks contribute to the prevalence of COVID-19 in the community, so workplaces could potentially amplify the rate of transmission. This is evident in the role lockdowns played in reducing transmission, with multiple workplaces being closed. The second impact is on the workforce. Workplace outbreaks result in potentially large proportions of the workforce being unwell and unable to work effectively [3]. This can have major financial impacts on both the company and the economy [4]. Therefore, understanding risk factors that contribute to workplace outbreaks is vital.

There are multiple factors that could affect the risk of workplace outbreaks. The frequency and size of outbreaks vary substantially by time and space, depending on the time-varying community prevalence of COVID-19 in the local population and the prevalence of workplaces and industries in the area. These are likely to be the main exposures contributing to outbreaks in the workplace, but there are other potential exposures that could amplify the risk of outbreaks occurring for given levels of COVID-19 prevalence. Demographic risk factors that have been associated with increased risk of infection include age, gender, ethnicity, and deprivation [5]. Workplace risk factors that are likely to influence outbreak risk include physical proximity in the workplace, contract stability of the workforce, and the land use characteristics of the local area.

Workplaces can present opportunities for a range of infectious diseases to spread due to conditions and contact duration, including respiratory syncytial virus (rsv) [6], influenza, and norovirus [7]. Infectious diseases can be introduced through providing services to infectious individuals, for example in the health and social care and in public facing work settings, and through workers carrying infectious diseases attending the workplaces. However, to study the influence of risk factors on the potential onward spread of infectious diseases requires disease surveillance data at levels of detail previously unavailable. For COVID-19 however, mass testing of both the community and the workforce has led to reliable metrics for both the rate of outbreaks in workplaces and the prevalence of infection in the community. Therefore, this provides a unique data setting to answer the question around risk factors for workplace COVID-19 outbreaks.

To mitigate the workplace impact of future waves of COVID-19, understanding the different risk factors is essential to ensure accurate and reliable health and safety messaging can be issued to workplaces. Therefore, using a combination of commuter, workplace, and population characteristics, our objective was to assess the importance of risk factors associated with workplace outbreak risk in different industry sectors. The findings of the study can support future pandemic preparedness.

## 2. Methods
2.1. Setting

This study examines variation in workplace COVID-19 cluster rates by Middle Layer Super Output Area (MSOA) in England according to workplace characteristics, and the characteristics of the local population of the workers and those of the area around the workplaces. We consider three time periods for the study: 20/06/2021 - 30/09/2021 (Delta variant dominant strain), 07/12/2021-20/02/2022 (Omicron variant dominant strain) [8], and 20/06/2021 - 20/02/2022 (i.e. overall, encompassing both periods and the intervening period). These study periods are considered since testing policy was relatively stable, with mass testing available in the community and many industries mandating staff testing strategies.

2.2. Data sources

We used population data to investigate potential risk factors for workplace clusters, overall and by sector. The data used in this study comes mainly from three sources: Office for National Statistics (ONS), Health and Safety Executive (HSE), and UK Health Security Agency (UKHSA). The ONS data used pertains to the demographic characteristics of workers and residents in England. The HSE data used pertains to the characteristics of workplaces in England. The UKHSA data used pertains to data on COVID-19, including testing data and vaccination records.

For this study, demographic characteristics have been obtained from the 2011 Census conducted by the ONS [9]. The variables extracted from the census were age, gender, ethnicity, and methods of commute to work. We also used data on indices of multiple deprivation (IMD) from the Department for Levelling Up, Housing and Communities [10].

Workplace COVID-19 clusters were identified primarily from the UKHSA Contact Tracing and Advisory Service (CTAS) data [11]. These data collected information on individuals who tested positive for COVID-19 in England and were contacted by the contact tracing service. For individuals who tested positive (and attended their workplaces 3 to 7 days prior to their positive test), their places of work were coded using the Unique Property Reference Numbers (UPRNs). This enabled workplace clusters of cases to be counted based on 2 or more cases of self-reported COVID-19 occurring at the same workplace building with a rolling 6-day period, using the method described in [12]. We use clusters as a proxy measure for workplace outbreaks in this analysis.

Mass testing data from UKHSA consisting of pillar 1 and pillar 2 tests was used to measure case-incidence over time [13]. During the study period, 20/06/2021 - 20/02/2022, testing policy was roughly constant across England. Therefore, infection incidence (the number of new infections each day) is assumed to be proportional to case-incidence (the number of positive tests each day) in this period, and case-incidence can be used as a proxy for the force of infection in the community. The final dataset used from UKHSA is from the National

Immunisation Management System (NIMS), which contains COVID-19 vaccination records [14]. These provide data on all vaccination doses given in England, from which we calculated the number and proportion of working age individuals vaccinated.

The National Population Database (NPD) [15] provides geographical estimates of population for five different categories of population: Residential, Sensitive, Workplace, Leisure and Transport. The NPD workplace layer is derived from the Inter-Departmental Business Register (IDBR) [16] and includes estimates of the number and location of workplaces and workers in GB by industry type using the Standard Industrial Classification (SIC). The number of workplaces in each Workplace Zone (a geographical area that is nested within an MSOA) were calculated by industry type, and subsequently aggregated to MSOA level. Workplaces with identified COVID-19 clusters were linked to the NPD workplace layer using business name, address and postcode information [12] so that a SIC code could be applied. Detailed level of workplace information was aggregated into 34 COVID-19 related sector groups [17], which were then aggregated into 11 broad industry sectors, to assist analysis and present the results.

Other workplace data used in this study includes data on physical proximity in the workplace at two-digit SIC grouping level [18] and the proportion of workers on permanent contracts at two-digit SIC grouping level [19]. Physical proximity represents the mean proximity score for each industry sector. The final workplace data used is the classification of each workplace MSOA by mobility class [20], which provides a refined definition of rural/urban classification.

## 2.3. Data preparation

Full details of data preparation are contained in the Supplementary material. The variables used in this study fall into 5 broad categories: workplace, case and cluster rates, commuter (characteristics of the workers who travel into workplaces), method of travel to work, residential (characteristics of people who live near the workplace). Table 1 shows the list of variables and groupings.

**Table 1:** *Variables considered in this study. Risk factor variables are the risk factors of interest in this study. Adjustment variables are included in the minimum adjustment set.*

| Variable | Type | Group |
|---|---|---|
| Physical proximity in the workplace | Risk factor | **Workplace** |
| Proportion of workers on permanent contracts | Risk factor | **Workplace** |
| Mobility class | Risk factor | **Workplace** |
| Proportion of commuters with 2 or more vaccination doses | Risk factor | **Commuter** |
| Proportion of commuters aged 18-29 | Risk factor | **Commuter** |
| Proportion of commuters aged 30-44 | Risk factor | **Commuter** |
| Proportion of commuters aged 45-59 | Risk factor | **Commuter** |
| Proportion of commuters aged 60-64 | Risk factor | **Commuter** |
| Proportion of commuters with an asian ethnicity | Risk factor | **Commuter** |
| Proportion of commuters with a black/african/caribbean ethnicity | Risk factor | **Commuter** |
| Proportion of commuters with a mixed/multiple/other ethnicity | Risk factor | **Commuter** |
| Proportion of commuters with a white ethnicity | Risk factor | **Commuter** |
| Proportion of commuters who identify as Female | Risk factor | **Commuter** |
| Proportion of commuters who identify as Male | Risk factor | **Commuter** |
| Commuter IMD quintile | Risk factor | **Commuter** |
| Proportion of commuters using bus/metro/tram | Risk factor | **Method of travel to work** |
| Proportion of commuters using taxi/vehicle passenger | Risk factor | **Method of travel to work** |
| Proportion of commuters using other transport | Risk factor | **Method of travel to work** |
| Proportion of commuters using train | Risk factor | **Method of travel to work** |
| Proportion of commuters using single occupancy | Risk factor | **Method of travel to work** |
| Workplace cluster rate | Adjustment | **Case and cluster rates** |
| Workplace cluster rate - other industries | Adjustment | **Case and cluster rates** |
| Commuter COVID-19 case rate | Adjustment | **Case and cluster rates** |
| Resident IMD quintile | Adjustment | **Residential** |
| Proportion of residents with 1 or more vaccination doses | Adjustment | **Residential** |
| Proportion of residents with 2 or more vaccination doses | Adjustment | **Residential** |
| Resident COVID-19 case rate | Adjustment | **Residential** |

2.4. Study size

We did not perform a sample size calculation when designing this study. Commuter characteristics, Work commute methods, and COVID-19 case rates and vaccination rates have coverage across all MSOAs. Industrial sector characteristics have coverage for all industry sectors considered. Of the total number of workplace outbreaks detected, 42% were excluded since these could not be reliably linked to a specific industry sector when merging CTAS data with NPD data.

2.5. Statistical methods

Our outcome of interest is the workplace COVID-19 cluster rate in the MSOA, which we define as the number of active clusters in workplaces in an area divided by the total number of workplaces in that area. Since the number of active clusters is an integer valued count, these can be considered as samples from a counting process, such as a Negative Binomial distribution. Therefore, instead of modelling the outbreak cluster rate directly, we modelled the time-varying number of active clusters in the workplace as the outcome variable of a negative binomial generalised linear model, and introduced the log transformed number of workplaces as an offset to transform this into a rate.

The risk factor variables that we were primarily interested in fall into three distinct groups: industrial sector characteristics, commuter characteristics, and work commute methods. These risk factors were selected by health and safety experts for examination in the current study. The full list of risk factors examined is shown in Table 1, alongside common adjustment variables.

For each risk factor, we fitted 3 models. The first included only the risk factor (Model 1: unadjusted). The second extended this by including the residential variables (characteristics of individuals living near the workplace), the log-transformed 7-day lagged values of COVID rates amongst commuters, the cluster rate in that industry, and the cluster rate in other industries (Model 2: minimal adjusted). Details on the adjustment variables can be found in Table 1. We take the logarithm of the lagged variables, which implies an approximately linear dependence on these variables. The third extended Model 2 by adding variables from other risk factor groups which are plausible causal antecedents of the risk factor of interest in the model, i.e. they do not occur later down the causal pathway (Model 3: priority risk factor-fully adjusted). Our adjustment sets were decided upon by discussion amongst the author team, which included consideration of plausible Directed Acyclic Graphs (DAGs) representing the relationships between the study variables. Acknowledging the complexity of the underlying process and the challenge of identifying a single DAG, we selected adjustment sets which would be compatible with the principles described by VanderWeele (2019) [21]. Further details on the construction of the statistical models and the potential bias are provided in the Supplementary Material.

Note that models 1 and 2 include the same explanatory variables for all risk factors. Model 3 uses different explanatory variables depending on which group of risk factors the factor of interest falls within. We opt to use a negative binomial generalised linear model rather than Poisson, as this better captures the error structure. When measuring the effect of vaccination, we consider the periods where the Delta and Omicron variants were dominant in the population under investigation independently, since vaccine uptake depends on time, and different COVID-19 variants have different levels of vaccine effectiveness [22].

Each industry is modelled independently, because different testing policies across industry sectors led to different case ascertainment rates. Therefore, clusters in one industry sector are not directly comparable to another industry sector, so a combined model looking at all industry sectors simultaneously would be invalid. For Education, we remove school holiday periods. We exclude Education outbreaks between 18/07/2021 - 12/09/2021, 24/10/2021 - 31/10/2021, 19/12/2021 - 09/01/2022. An extra week is excluded after each holiday period to account for delays from infection to cases being detected which adds a lag before workplace clusters are detected after returning from the holidays. For the remaining industry sectors, the full study period is included.

## 3. Results
### 3.1. Descriptive results
The total number of workplaces included in this study is given by the number of distinct workplaces in the NPD. Across the study period, 1,149,007 workplaces are included, with the industry breakdown given in Table 2 The study period runs from 20/06/2021 to 20/02/2022, which leads to 36 weeks being modelled, and we have 6721 distinct MSOAs. We have 11 industry sectors that are considered. Altogether, we have 801,874 data points, where each data point corresponds to a unique combination of MSOA, week, and industry. Note that not all permutations are included, as for some MSOAs not all industry sectors are present.

Table 2 shows a descriptive summary of the data used in this study by industry. The mean is calculated based on the number of workplaces in each industry, rather than the number of employees, since the modelling is done at workplace-level. Figure 1 shows the cluster rate (number of active clusters in the workplace / number of workplaces) at a national level over time.

*Table 2: Descriptive summaries of the data used in the study. Proportions are recorded as percentages.*

| Variable | Value by industry | | | | | | | | | | |
|---|---|---|---|---|---|---|---|---|---|---|---|
| | Services | Utilities | Education | Transport, distribution and warehousing | Mining and Quarrying | Manufacturing | Public service activities | Construction | Human health and social work | Waste management and remediation | Agriculture, forestry and fishing |
| **Workplace characteristics** | | | | | | | | | | | |
| Total number of workplaces included | 800710 | 3624 | 24636 | 77771 | 108 | 68437 | 7102 | 83224 | 68449 | 1202 | 13744 |
| Mean physical proximity in the workplace | 58 | 53 | 68 | 59 | 55 | 54 | 60 | 61 | 74 | 54 | 48 |
| Mean proportion of workers on permanent contracts | 94% | 98% | 91% | 96% | 99% | 96% | 96% | 97% | 94% | 95% | 93% |
| Most common mobility class | Exurban | Exurban | Exurban | Exurban | Exurban | Suburban | Exurban | Exurban | Exurban | Suburban | Suburban |
| **Case and cluster rates** | | | | | | | | | | | |
| Mean number of active clusters | 0.54 | 0.14 | 0.48 | 0.25 | 0.11 | 0.28 | 0.27 | 0.11 | 0.24 | 0.11 | 0.09 |
| Mean log(commuter_case_rate_lag_7day) | -7.61 | -7.60 | -7.70 | -7.61 | -7.62 | -7.60 | -7.61 | -7.61 | -7.61 | -7.60 | -7.65 |
| Mean log(cluster_rate_other) | -16.02 | -11.25 | -16.93 | -13.16 | -14.55 | -13.84 | -12.32 | -12.74 | -15.07 | -11.90 | -14.82 |
| Mean log(cluster_rate_lag_7day) | -19.24 | -22.27 | -17.64 | -21.11 | -22.70 | -20.46 | -19.89 | -23.23 | -21.04 | -22.82 | -23.50 |
| **Commuter characteristics** | | | | | | | | | | | |
| Mean proportion of commuters with 2 or more vaccination doses | 78.3% | 78.5% | 79.1% | 78.2% | 80.8% | 78.6% | 78.2% | 78.7% | 78.4% | 78.5% | 81.2% |
| Mean proportion of commuters aged 18-29 | 27.4% | 25.1% | 20.5% | 23.0% | 21.3% | 21.5% | 20.2% | 22.2% | 21.8% | 21.4% | 19.4% |
| Mean proportion of commuters aged 30-44 | 36.9% | 41.2% | 36.3% | 37.5% | 36.2% | 36.0% | 37.4% | 36.6% | 36.0% | 35.6% | 31.8% |
| Mean proportion of commuters aged 45-59 | 30.1% | 29.1% | 36.2% | 33.3% | 34.5% | 35.5% | 36.4% | 34.3% | 35.4% | 35.8% | 39.0% |
| Mean proportion of commuters aged 60-64 | 5.6% | 4.6% | 7.0% | 6.3% | 8.1% | 7.1% | 6.1% | 6.9% | 6.8% | 7.2% | 9.8% |
| Mean proportion of commuters with an asian ethnicity | 7.7% | 8.2% | 6.1% | 7.2% | 3.9% | 5.4% | 5.3% | 5.7% | 7.2% | 4.0% | 1.2% |
| Mean proportion of commuters with a black/african/caribbean ethnicity | 3.4% | 3.6% | 2.8% | 2.8% | 1.6% | 1.7% | 2.8% | 2.5% | 3.4% | 1.8% | 0.4% |
| Mean proportion of commuters with a mixed/multiple/other ethnicity | 2.9% | 3.0% | 2.2% | 2.3% | 1.7% | 1.6% | 1.9% | 2.0% | 2.3% | 1.5% | 0.8% |
| Mean proportion of commuters with a white ethnicity | 86.0% | 85.2% | 89.0% | 87.7% | 92.7% | 91.3% | 90.0% | 89.8% | 87.1% | 92.8% | 97.6% |
| Mean proportion of commuters using bus/metro/tram | 17.4% | 21.5% | 10.2% | 10.6% | 11.5% | 6.6% | 11.4% | 9.7% | 11.6% | 5.2% | 2.3% |
| Mean proportion of commuters using taxi/vehicle passenger | 5.4% | 4.0% | 5.7% | 6.3% | 5.4% | 7.2% | 5.2% | 6.1% | 5.9% | 7.3% | 7.4% |
| Mean proportion of commuters using other transport | 0.4% | 0.3% | 0.5% | 0.4% | 0.6% | 0.4% | 0.4% | 0.5% | 0.5% | 0.4% | 0.8% |
| Mean proportion of commuters using train | 9.1% | 18.0% | 4.4% | 5.3% | 8.2% | 2.4% | 6.6% | 5.1% | 4.4% | 1.7% | 0.8% |
| Mean proportion of commuters using single occupancy | 67.6% | 56.3% | 79.2% | 77.4% | 74.3% | 83.4% | 76.4% | 78.7% | 77.6% | 85.5% | 88.7% |
| Mean proportion of commuters who identify as Female | 48.0% | 39.4% | 52.8% | 36.8% | 35.6% | 34.4% | 50.5% | 41.0% | 54.4% | 30.8% | 40.9% |
| Mean proportion of commuters who identify as Male | 52.0% | 60.6% | 47.2% | 63.2% | 64.4% | 65.6% | 49.5% | 59.0% | 45.7% | 69.3% | 59.1% |
| Most common IMD quintile among commuters | 3 | 3 | 3 | 3 | 3 | 3 | 3 | 3 | 3 | 3 | 3 |
| **Residential characteristics** | | | | | | | | | | | |
| Mean proportion of residents with 2 or more vaccination doses | 77.9% | 75.4% | 78.4% | 76.5% | 82.9% | 77.7% | 73.5% | 77.3% | 77.6% | 76.3% | 84.5% |
| Most common IMD quintile among residents | 3 | 2 | 3 | 2 | 3 | 3 | 2 | 3 | 3 | 2 | 3 |

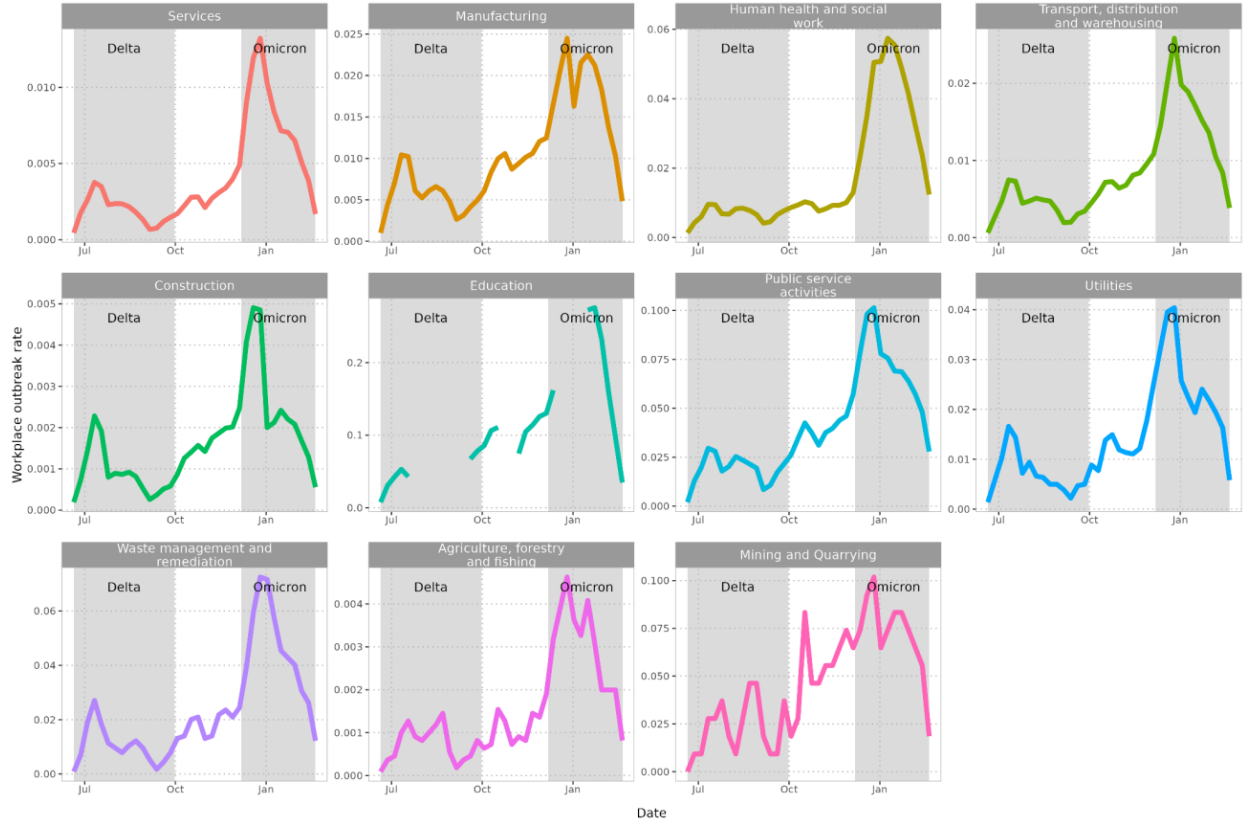

**Figure 1:** *Workplace cluster rates (total number of active clusters in the workplace divided by total number of workplaces) over time, broken down by industry sector. The shaded regions mark the Delta and Omicron periods. The full time series is the overall study period.*

### 3.2. Main results

The model identifies that many of the risk factors are associated with the risk of COVID-19 clusters in the workplace. Some risk factors have a clear trend that is apparent across industries. Other risk factors differ substantially across industries, with the direction and magnitude of relationships changing significantly. The results discussed and shown in Tables 3a and 3b are for the priority risk factor fully adjusted model. Supplementary Tables S1 and S2 show the results for the unadjusted and minimal adjusted models for all sectors, respectively. Comparing results across the unadjusted, minimal adjusted, and the priority risk factor fully adjusted model, some of the results are substantially different, demonstrating that the identified causal structure has a large influence, with correlations and associations between the included variables leading to different effects. Table 3a contains results for the 5 industry sectors identified as high importance by HSE: Education; Human health and social work activities; Services; Transport, Distribution and Warehousing; and Manufacturing. Table 3b contains results for the remaining 6 industry sectors. The effect of vaccination is considered separately for the Delta and Omicron periods, due to different vaccine effectiveness across different variants. For all other risk factors, we consider the full study period.

*3.2.1. Workplace characteristics*

Across all industries where sufficient data are available, increasing physical proximity in the workplace was associated with increased risk of outbreaks, except for Construction where the effect was not statistically significant. Increasing the proportion of workers on permanent contracts was associated with reduced risk in all industries except Construction. In Construction, the effect was an increase in risk. The relationship with mobility class varied substantially across the industries, but was generally found to be a significant risk factor.

*3.2.2. Commuter vaccination*

Increasing the cumulative proportion of commuters that have received 2 or more doses of vaccination against Sars-CoV-2, during the Delta and Omicron periods, was associated with reduced risk (or to have no significant effect) for all industries except Education. In Education, vaccination was associated with increased risk during Delta and no significant effect during Omicron. However, this is likely driven by the lack of vaccination in students [14], who are the main exposure for individuals working in education. Therefore, since prevalence of Delta was growing exponentially up to the holiday period (which makes up the majority of the study period in the Education sector), and any potential effectiveness of vaccination may be small due to the lack of vaccination in students, the model is confounding increased vaccine uptake (which increases monotonically with time) with increased transmission risk, rather than associating increased Delta prevalence with increased transmission risk.

*3.2.3. Demographic characteristics*

Many of the demographic characteristics were significant risk factors. Across every industry, having higher proportions of individuals in the 60-64 age group was associated with substantially reduced risk of clusters (or no significant effect). In all industries, a higher proportion of individuals in either the 18-29 age group or the 44-59 age group was associated with an elevated risk. In all industries, the presence of higher proportions of individuals from black/African/Caribbean ethnicities were associated with increased cluster risk (or no significant effect) [23]. Mixed/multiple/other were associated with reduced risk in all industries except Public Service Activities and Agriculture, Forestry, and Fishing. The trend across the remaining groups of ethnicities varied across industries. The association with sex varied across industries, with 4 industries seeing increased risk from a higher proportion of female workers and 6 industries seeing reduced risk. In all industries, increasing the IMD quintile (reducing deprivation) was associated with reduced risk of workplace outbreaks, except for Mining and quarrying and Agriculture, Forestry and Fishing, where the effect was not significant.

*3.2.4. Methods of commute to work*

The methods of transport that commuters used to travel to work generally had a significant association with the risk of workplace outbreaks across all industries, though the association varied substantially across industries. Taxi (or vehicle passenger) was associated with increased risk (or no effect) across all industries except Education, Utilities, and Public service activities. Trains were associated with reduced risk (or no effect) in all industries. The effect with bus, tram or metro use varied, with some industries seeing increased risk and some reduced risk. Some of the effect of public transport could be due to repeat infections over time on public transport leading a higher level of natural immunity at the time of the study.

3.3. High importance industry sectors

Here we provide a breakdown of the results for the high importance industry sectors: Services; Education; Transport, Distribution and Warehousing; Manufacturing; Human health and Social Work (Table 3a). We focus on the risk factors that are highly variable across industry sectors.

In Education, the proportion of commuters in the mixed/multiple/other group of ethnicities was associated with the lowest risk followed by asian, then white, with black/african/caribbean associated with the highest risk. Rural areas are associated with substantially lower risk than others [24], followed by Exurban, then Metropolitan, and finally Suburban. Sub-industries with a higher proportion of staff on permanent contracts were associated with increased risk. A higher proportion of female commuters was associated with increased risk.

In Transport, distribution and warehousing, the proportion of commuters in the mixed/multiple/other group of ethnicities was associated with the lowest risk followed by asian, then white, with black/african/caribbean ethnicities associated with the highest risk. Rural and Metropolitan areas were associated with reduced risk, with Exurban and Suburban having similar risks. Sub-industries with a higher proportion of staff on permanent contracts were associated with reduced risk. A higher proportion of female commuters was associated with increased risk.

In Services, the proportion of commuters in the mixed/multiple/other group of ethnicities was associated with the lowest risk followed by asian and white with similar risk, with black/african/caribbean associated with the highest risk. Exurban areas were associated with the lowest risk, followed by Metropolitan and Rural with similar risk, and finally Suburban. Sub-industries with a higher proportion of staff on permanent contracts were associated with reduced risk. A higher proportion of female commuters was associated with reduced risk.

In Manufacturing, the proportion of commuters in the mixed/multiple/other group of ethnicities was associated with the lowest risk followed by asian, then white, with black/african/caribbean ethnicities associated with the highest risk. Metropolitan areas were associated with the lowest risk, with Rural, Exurban, and Suburban having comparable risks to each other. Sub-industries with a higher proportion of staff on permanent contracts were associated with reduced risk. A higher proportion of female commuters was associated with reduced risk.

In Human health and social care, the proportion of commuters in the mixed/multiple/other group of ethnicities was associated with the lowest risk followed by white, then asian, with black/african/caribbean associated with the highest risk. Exurban areas are associated with the lowest risk, followed by Metropolitan and Suburban areas, with similar risk to each other, and finally Rural with the highest risk. Sub-industries with a higher proportion of staff on permanent contracts were associated with increased risk. A higher proportion of female commuters was associated with increased risk.

*Table 3a:* HSE priority-specific fully adjusted model, percentage change in risk of workplace outbreak cluster rate by industry. Services; Education; Transport, distribution and warehousing; Manufacturing; Human health and social work. Confidence intervals are 90%.

| Variable | Percentage change in risk by industry (fully adjusted model) | | | | |
|---|---|---|---|---|---|
| | Services | Education | Transport, distribution and warehousing | Manufacturing | Human health and social work |
| **Workplace - proximity** | | | | | |
| *Physical proximity in the workplace* | 5.3 (5.1,5.5) | NA | 12 (12,13) | 4.6 (4.2,5.1) | NA |
| **Workplace - permanence** | | | | | |
| *Proportion of workers on permanent contracts* | -0.77 (-1.3,-0.25) | NA | -22 (-24,-21) | -13 (-14,-12) | NA |
| **Workplace - mobility class** | | | | | |
| *Mobility class - Exurban* | -17 (-19,-16) | -9.5 (-11,-7.7) | 0.18 (-3.3,3.8) | -1.5 (-4.1,1.3) | -3.6 (-5.9,-1.3) |
| *Mobility class - Metropolitan* | -9.3 (-11,-7.9) | -2.2 (-3.9,-0.44) | -7.7 (-10,-4.8) | -7.1 (-10,-4.1) | -0.44 (-2.4,1.6) |
| *Mobility class - Rural* | -11 (-14,-8.4) | -44 (-46,-42) | -6.3 (-11,-1.1) | -0.16 (-4.3,4.2) | 11 (6.7,16) |
| *Mobility class - Suburban* | baseline | baseline | baseline | baseline | baseline |
| **Commuter - vaccination** | | | | | |
| *Proportion of commuters with 2 or more vaccination doses - Delta* | -3.7 (-4,-3.5) | 1.8 (1.4,2.3) | -3.2 (-3.8,-2.7) | -1.7 (-2.2,-1.1) | -1.8 (-2.3,-1.3) |
| *Proportion of commuters with 2 or more vaccination doses - Omicron* | -3.2 (-3.5,-2.9) | 0.24 (-0.12,0.61) | -4.2 (-4.8,-3.7) | -1.3 (-1.9,-0.79) | -3.2 (-3.5,-2.8) |
| **Commuter - age** | | | | | |
| *Proportion of commuters aged 18-29* | -0.83 (-1,-0.62) | 0.36 (0.081,0.64) | 2.7 (2.2,3.2) | -0.78 (-1.3,-0.3) | -1.1 (-1.5,-0.77) |
| *Proportion of commuters aged 30-44* | baseline | baseline | baseline | baseline | baseline |
| *Proportion of commuters aged 45-59* | 3.4 (3.1,3.7) | 2.5 (2.2,2.7) | 5.1 (4.5,5.6) | 3.7 (3.2,4.1) | 2.2 (1.9,2.5) |
| *Proportion of commuters aged 60-64* | -4.6 (-5.2,-4.1) | -1.1 (-1.6,-0.61) | -2.9 (-3.8,-1.9) | -12 (-13,-11) | -3.2 (-3.8,-2.6) |
| **Commuter - ethnicity** | | | | | |
| *Proportion of commuters with an asian ethnicity* | 0.046 (-0.065,0.16) | -0.25 (-0.37,-0.12) | -1.2 (-1.4,-0.94) | -0.48 (-0.7,-0.26) | 0.21 (0.058,0.37) |
| *Proportion of commuters with a black/african/caribbean ethnicity* | 1.6 (1.3,1.9) | 0.49 (0.19,0.79) | 2.7 (2.1,3.4) | 1.8 (0.9,2.7) | 2.4 (2.1,2.7) |
| *Proportion of commuters with a mixed/multiple/other ethnicity* | -12 (-12,-11) | -4.9 (-5.7,-4.2) | -6.4 (-7.9,-4.8) | -9.2 (-11,-7.5) | -9.4 (-10,-8.5) |
| *Proportion of commuters with a white ethnicity* | baseline | baseline | baseline | baseline | baseline |
| **Commuter - sex** | | | | | |
| *Proportion of commuters who identify as Female* | -1.3 (-1.4,-1.3) | 1.4 (1.3,1.5) | 1 (0.88,1.2) | -1.1 (-1.3,-0.96) | 0.31 (0.2,0.42) |
| *Proportion of commuters who identify as Male* | baseline | baseline | baseline | baseline | baseline |
| **Commuter - IMD** | | | | | |
| *Commuter IMD quintile* | -15 (-16,-14) | -5.2 (-6.4,-4) | -14 (-16,-12) | -10 (-12,-8.3) | -17 (-18,-16) |
| **Method of travel to work** | | | | | |
| *Proportion of commuters using bus/metro/tram* | -0.0093 (-0.13,0.12) | -1.4 (-1.6,-1.3) | 1.4 (1.1,1.6) | 0.14 (-0.22,0.49) | 0.049 (-0.12,0.22) |
| *Proportion of commuters using other transport* | 20 (18,22) | 0.56 (-1.2,2.3) | 8.7 (5.7,12) | 2.6 (-1.4,6.7) | 17 (14,19) |
| *Proportion of commuters using single occupancy* | baseline | baseline | baseline | baseline | baseline |
| *Proportion of commuters using taxi/vehicle passenger* | 5.3 (4.8,5.7) | -1.9 (-2.3,-1.4) | 4.1 (3.5,4.7) | 1.6 (1.1,2.1) | 4.3 (3.8,4.9) |
| *Proportion of commuters using train* | -1.1 (-1.3,-0.99) | -2.4 (-2.6,-2.2) | -2.6 (-2.8,-2.3) | -4.2 (-4.8,-3.7) | -2.4 (-2.7,-2.2) |

*Table 3b:* HSE priority-specific fully adjusted model, percentage change in risk of workplace outbreak cluster rate by industry. Utilities; Mining and quarrying; Public service activities; Construction; Waste management and remediation; Agriculture, forestry, and fishing.

*Confidence intervals are 90%.*

| | Percentage change in risk by industry (fully adjusted model) | | | | | |
|---|---|---|---|---|---|---|
| Variable | Utilities | Mining and Quarrying | Public service activities | Construction | Waste management and remediation | Agriculture, forestry and fishing |
| **Workplace - proximity** | | | | | | |
| Physical proximity in the workplace | 18 (14,22) | NA | NA | -0.3 (-2,1.4) | NA | NA |
| **Workplace - permanence** | | | | | | |
| Proportion of workers on permanent contracts | -16 (-21,-12) | NA | NA | 1e+02 (60,1.6e+02) | NA | NA |
| **Workplace - mobility class** | | | | | | |
| Mobility class - Exurban | 15 (0.27,32) | 8.5 (-25,57) | -34 (-38,-31) | -14 (-21,-7.8) | 10 (-5.7,29) | -39 (-49,-27) |
| Mobility class - Metropolitan | -52 (-58,-47) | -82 (-91,-64) | 4.3 (0.29,8.4) | -20 (-25,-15) | 42 (23,64) | 1e+02 (26,2.3e+02) |
| Mobility class - Rural | -14 (-30,5.4) | 24 (-16,83) | -20 (-26,-13) | -20 (-29,-9.3) | 46 (19,79) | -69 (-75,-63) |
| Mobility class - Suburban | baseline | baseline | baseline | baseline | baseline | baseline |
| **Commuter - vaccination** | | | | | | |
| Proportion of commuters with 2 or more vaccination doses - Delta | -2.4 (-4.4,-0.41) | 6.8 (-7.2,23) | -3.2 (-4,-2.3) | -1.5 (-2.9,-0.21) | -2.2 (-5.3,0.98) | -7 (-12,-2) |
| Proportion of commuters with 2 or more vaccination doses - Omicron | -8.4 (-10,-6.3) | 5.2 (-3,14) | -6 (-6.7,-5.2) | -1.5 (-3,0.011) | 1.4 (-1.3,4.1) | -9.9 (-14,-6.1) |
| **Commuter - age** | | | | | | |
| Proportion of commuters aged 18-29 | 5.3 (4,6.6) | 0.86 (-3.5,5.4) | 1.1 (0.54,1.7) | 1.3 (0.15,2.5) | 5.6 (3.3,7.9) | 3.5 (0.84,6.3) |
| Proportion of commuters aged 30-44 | baseline | baseline | baseline | baseline | baseline | baseline |
| Proportion of commuters aged 45-59 | 8.8 (7.4,10) | 10 (4.8,17) | -0.16 (-0.78,0.45) | 4 (2.8,5.1) | 6.6 (4.4,8.9) | -2 (-5.1,1.2) |
| Proportion of commuters aged 60-64 | 0.28 (-2.5,3.2) | -12 (-18,-6.5) | -2.3 (-3.6,-1) | -3.7 (-5.9,-1.4) | -0.079 (-3.5,3.5) | -8.3 (-13,-3.4) |
| **Commuter - ethnicity** | | | | | | |
| Proportion of commuters with an asian ethnicity | 1.6 (0.41,2.7) | 9.5 (3.6,16) | 0.12 (-0.2,0.45) | 1.2 (0.55,1.8) | 1.4 (0.16,2.6) | 2.1 (-0.37,4.6) |
| Proportion of commuters with a black/african/caribbean ethnicity | 8.9 (6.8,11) | 25 (-7,67) | 0.47 (-0.069,1) | 0.74 (-0.6,2.1) | 3.5 (0.31,6.7) | -1.9 (-12,9.7) |
| Proportion of commuters with a mixed/multiple/other ethnicity | -30 (-34,-25) | -43 (-63,-14) | 6.4 (4.4,8.5) | -11 (-14,-8) | -9.7 (-16,-3) | 55 (32,82) |
| Proportion of commuters with a white ethnicity | baseline | baseline | baseline | baseline | baseline | baseline |
| **Commuter - sex** | | | | | | |
| Proportion of commuters who identify as Female | -0.56 (-0.96,-0.17) | -2.1 (-3.2,-0.91) | -1.2 (-1.4,-1) | -1 (-1.4,-0.62) | 1.3 (0.72,1.9) | 0.81 (-0.54,2.2) |
| Proportion of commuters who identify as Male | baseline | baseline | baseline | baseline | baseline | baseline |
| **Commuter - IMD** | | | | | | |
| Commuter IMD quintile | -31 (-36,-26) | 23 (-6.8,61) | -14 (-17,-11) | -19 (-23,-14) | -11 (-20,-1.1) | -3.5 (-15,10) |
| **Method of travel to work** | | | | | | |
| Proportion of commuters using bus/metro/tram | -1.7 (-2.5,-0.81) | 2.7 (-6.2,12) | -2.2 (-2.5,-1.8) | -1.1 (-1.7,-0.49) | 4.1 (1.8,6.3) | 0.62 (-2.5,3.9) |
| Proportion of commuters using other transport | 13 (-2.1,29) | -32 (-49,-8.1) | 1.7 (-2.7,6.3) | 3.8 (-6.4,15) | 8.2 (-8.5,28) | -18 (-32,-1.8) |
| Proportion of commuters using single occupancy | baseline | baseline | baseline | baseline | baseline | baseline |
| Proportion of commuters using taxi/vehicle passenger | -2.4 (-4.4,-0.36) | -5.6 (-13,2.8) | -2.4 (-3.5,-1.4) | 3.1 (1.2,4.9) | -1.4 (-4.1,1.4) | 4.1 (1.6,6.6) |
| Proportion of commuters using train | -5.6 (-6.4,-4.9) | -2.1 (-12,8.6) | -1.5 (-1.8,-1.3) | 0.29 (-0.23,0.82) | 0.54 (-2,3.2) | -0.42 (-5.3,4.8) |

## 4. Discussion

### 4.1. Key results

We set out to quantify the risk factors contributing to workplace clusters of COVID-19 in different industry settings. We found in most industries increased physical proximity in the workplace increased the risk of workplace clusters, and increased vaccination uptake reduced the risk of workplace clusters. Such results were expected due to known vaccine effectiveness against COVID-19 [25] and the airborne transmission being highly dependent on physical proximity. Other workplace demographics, such as the proportion of workers on permanent contracts and the mobility class of the local MSOA were found to have substantial variation across industries. Commuter demographic risk factors also varied across industry, but for most industries, a higher proportion of black ethnicities, 45-59 year olds, living in deprived areas (areas with reduced IMD), was associated with increased outbreak risk, and a higher proportion of 60-64 years in the workforce were associated with reduced cluster risk in the majority of industries. This may be related to the extra protection policy in the workplace for older and vulnerable workers during the pandemic. Workplaces where a higher proportion of commuters were expected to use Taxis (or travel as a passenger in a private car) and other forms of transport were found to have increased cluster risk for many industries, with trains, single occupancy, or buses, trams and metro typically associated with the lowest risk.

This work relied on linking local individual, workplace and community data on a population level. Through collaboration with workplace health and safety regulators (HSE), population statistics authorities (ONS), and healthcare data authorities (UKHSA), we were able to link relevant population data from multiple sources at the national level to identify and investigate workplaces with potentially increased risk of COVID-19 clusters. The new ways of working and the innovative approach of using data linkages to investigate multiple and complex social, environmental and behavioural determinants associated with where people live and work can be applied to support future pandemic preparedness and wider occupational and environmental health research more generally.

4.2. Limitations

This analysis was conducted using multiple data sources linked by geographical unit (MSOA) with, on average, 2532 workers per MSOA included in our study. Ideally, such analysis would be conducted using data at an individual-or workplace-level, but such data are not generally readily available at the national level. Therefore, the commuter demographics had to be approximated, using the expected demographics of commuters in the MSOA where people live and industry distribution in the MSOA where people work. If some workplaces have very atypical commuter demographics, this would not be captured by the data. Additionally, because the latest 2021 Census was not yet available for this analysis, the 2011 Census was the closest data that could be used to capture the demographic variables. If the data for some locations have seen substantial changes since 2011, this may not accurately reflect the conditions in which people live and work for the workplace COVID-19 clusters analysis period. The nature of the data creates the possibility of ecological bias, whereby relationships observed at population-level differ from those at the level of individuals or workplaces. Further potential biases, and our approaches for reducing these, are described in the Supplementary Material.

Motivated by likely causal pathways considered by using DAGs, we have adjusted for key confounders in these pathways. However, we have not considered causal relationships between variables in the same group (commuter characteristics; workplace characteristics; methods of work commute). Therefore, causal relationships between demographic variables will not be captured. Moreover, the challenge of creating a DAG capable of capturing the complex network of factors relating to this question is high. We fitted a series of models for each candidate risk factor in order to investigate the impact of different adjustment sets on the estimated relationships.

From the DAG, we identified workplace ascertainment rate as an important factor. However, this variable is unobserved. Since workplace ascertainment rate varies highly across industry sectors, due to different testing policies, this means that we were unable to compare risk factors across industry sectors in this study.

One of the major covariates adjusted for in this analysis is the expected COVID-19 case rate among commuters. This is vital to control for, since some outbreaks will be expected if community COVID-19 case rates where people live are high. We have assumed that the ascertainment rate of workplace COVID-19 cases depends only on the type of workplace and not the location. There may be some variation due to local test availability which is not captured here.

Data on workplace physical proximity and the proportion of employees (permanence) is only available at two-digit SIC grouping level. Therefore, in this analysis, we assume that all workplaces in a given two-digit SIC grouping have the same proportional distribution of physical proximity and permanence of employment contract. This would not allow more detailed analysis for individual workplaces. Therefore, the estimated effect size reflects the risk in the different sub-industries in each industry sector, rather than individual workplaces. For some industry sectors (Agriculture, Forestry and Fishing, Human Health and Social Work, Mining and Quarrying, Public service activities, and Waste Management and Remediation), there are too few two-digit SIC groupings, so physical proximity and permanence are not considered for these industries.

### 4.3. Interpretation
This work provides a measure of how different risk factors contributed to the risk of COVID-19 clusters in workplaces for different industry settings. This study was performed at a population level, so does not reflect individual level risk. The commuter characteristic and method of travel risk factors show how the expected characteristics of the local workforce are associated with the risk of workplace clusters, rather than the risk of being involved in a cluster for individuals according to their characteristics. The workplace characteristics reflect how the characteristics of sub-industries within the industry sector are associated with the risk of workplace cluster occurrence. The mobility class characteristics reflect how the local area where the workplace was located affects the risk of workplace clusters.

### 4.4. Generalisability
Although this study has focused on COVID-19, the methods used may be generalisable to other communicable and non-communicable diseases that frequently lead to workplace clusters. The range of potential risk factors identified, and methods for linking these data at a population level, are likely to be relevant for most diseases. However, the influence of each risk factor may not be generalisable, since these will depend on the mode of transmission and the susceptibility profiles of different demographic groups. A challenge in applying these methods to other diseases may be data availability on the pathogen, or pathogens, of interest. The wealth of community surveillance data on COVID-19 led to reliable estimates of the background force of infection and vaccination rates, which were important to adjust for in the statistical analysis. In the absence of such data, back-calculation methods or cohort surveillance studies may be needed to apply this methodology to other pathogens.

**Author contributions**

| |
|---|
| Conceptualization: CO, RA, TW, OD, MG, JW, TA, YC |
| Literature review: RA, MG, JW, TA, YC |
| Data Curation: CO, RC, RP, JD, OJ, TB, CWH |
| Formal Analysis: CO, RP, MT |
| Methodology: CO, RP, OJ, MT, MG, JW, OD |
| Software: CO, RP, MT, OJ, CWH, TB, RC, JD |
| Visualization: CO |
| Writing – Original Draft Preparation: CO, OD, RA, JW, MG, TA, YC |
| Writing – Review & Editing: CO, OD, RA, JW, MG, TA, YC, TW |

**Funding statement:**
This work was funded by the PROTECT (Partnership for Research in Occupational, Transport and Environmental COVID Transmission) COVID-19 National Core Study on Transmission and Environment and managed by the Health and Safety Executive on behalf of HM Government.

**Conflict of Interest**
The authors declare that they have no conflict of interest.

**Disclosure Statement:** The contents of this paper, including any opinions and/or conclusions expressed, are those of the authors alone and do not necessarily reflect Health and Safety Executive policy or UK Health Security Agency or Office the National Statistics views.

**Ethics statements**
This study was conducted as part of the PROTECT National Core Study. Work was undertaken in line with national data regulations.

**Data availability statement**
The data has not been made available due to the sensitivity of workplace-level data. An application for data access can be made to the UK Health Security Agency. Data requests can be made to the Office for Data Release (https://www.gov.uk/government/publications/accessing-ukhsa-protected-data/accessing-ukhsa-protected-data) and by contacting DataAccess@ukhsa.gov.uk. All requests to access data are reviewed by the Office for Data Release and are subject to strict confidentiality provisions in line with the requirements of: the common law duty of confidentiality, data protection legislation (including the General Data Protection Regulation), Caldicott principles, the Information Commissioner's statutory data sharing code of practice, and the national data opt-out programme.
Access to the National Population Database can be requested by contacting: npd@hse.gov.uk.

# Supplementary Material

## S.1. Supplementary methods
S.1.1. Data preparation
*S.1.1.1. Flow to work*
Many of the variables in the datasets describe the residential characteristics of workers. To obtain commuter (that is, a worker, who travels from home to work) demographics, we need to understand where these individuals work. To do this, we use the flow to work data from the 2011 census. For each residential MSOA of where the workers lived, we calculate the proportion of workers going to each workzone in England. This provides a network describing where individuals are expected to work.

*S.1.1.2. Industry distributions*
Analysis was performed with MSOA as the unit of analysis. Therefore, we aggregated the commuter data from workzone level to MSOA level. To approximate the distribution of commuters across industries in each MSOA, we weighted the workzone level variables by the industry distribution across that workzone. This led to the assumption that within an individual workzone, commuters are uniformly distributed across the industries, based on the proportion of each industry present. At MSOA level, this led to the commuter characteristics being weighted by the industry distributions in the workzones to which these individuals commute.

*S.1.1.3. COVID-19 commuter case rates*
To understand background risk, we quantified the incoming infection risk for each workplace, which is a function of community prevalence. We did this by converting the community COVID-19 case rates into a commuter COVID-19 case rate. We began by calculating the case rate for each MSOA through time; we geographically matched Pillar 2 SARS-CoV-2 PCR tests to MSOA by postcode, removing travel associated tests. Only symptomatic tests were modelled, to avoid the confounder of different asymptomatic screening policies of different workplaces. These tests were relatively sparse at the MSOA level, so we smoothed case rates using a generalised additive model (GAM) with a quasi-poisson error structure and log-link. The number of positive tests, n, is smoothed for each MSOA, using a separate cubic regressions spline for each MSOA, $f(t,m)$. We expect that the case rates of neighbouring MSOAs would be more closely correlated than those that do not share a border. We therefore fit a Markov random field term that adjusts MSOA intercepts ($\propto_m$) based on its neighbours. We label this term $g(m)$. In full, this gives the expression:

$$\log(\lambda_{t,m}) = \propto_m + \beta_d + g(m) + f(t,m) + \log(P)$$

We used a log-offset term in the model to scale the number of tests by the MSOA population size ($P$), to give a per-capita rate. A day of week effect controlled for weekly cycles in test reporting and was modelled as a random effect. By modelling the day of week effect in this way, we could omit it from the modelled case rate giving only the central trend. This allows us to better describe the true rate, as opposed to weekly anomalies in reporting and testing. Only people aged 18-64 were included in the population offset term and testing data to better reflect the working age population per MSOA. To save on computational time, we analysed MSOA case rates at the local authority level; the data were subset to specific lower-tier local authority (LTLA) and the model described above fitted for all time points within the geography.

To calculate an incoming force of infection into each sector in each workzone MSOA, we began by averaging MSOA case rate across each week of the study period ($\bar{\lambda}_m$). For each workzone, there was a measure of how many people commuted from their resident MSOA into that workzone. For each workzone with $N_{m,w}$ commuters from $M_w$ resident MSOAs to the workzone:

$$C_w = \frac{\sum_{m=1}^{M_w} \bar{\lambda}_m \times N_{m,w}}{\sum_{m=1}^{M_w} N_{m,w}}.$$

Here, $C_w$ is the estimated number of infected commuters out of the total number of commuters coming into the workzone. For each workzone, we knew the proportion of businesses in workzone belonging to each "industry" ($\rho_{w,i}$). We assume that these commuters uniformly attended each industry, so to calculate a per-industry, per-workzone per-capita case rate we scaled $C_w$ by $\rho_{w,i}$.

*S.1.1.4. COVID-19 vaccination*
To provide a measure of COVID-19 vaccination levels, we counted the expected proportion of commuters into each workplace who have had two or more vaccination doses. We also considered the counts of the number of residents in each workplace MSOA who have had two or more vaccination doses.

*S.1.1.5. COVID-19 clusters*
We were interested in the number of active clusters in each MSOA, broken down by industry. To calculate this, we first define an "event" as an occasion where a COVID-19 case self-reported as being present at a location/potential site of infection, recorded via the backwards contact-tracing process. A "cluster" is then defined as a collection of events which occurred at a single location (common exposure event), defined by the Unique Property Reference Number (UPRN) as a spatial reference identifier associated with the event in the CTAS data; and where each event occurs within some episode period (in our case, defined as six full days) of the previous event. A cluster is "active" on a given date if that date lies between the dates of the chronologically first and last events which belong to the cluster, inclusive of those dates.

As a persistent identifier, the UPRN was additionally used to attach industry information to each cluster. For each UPRN, full address information was extracted from the subset of OS AddressBase Plus corresponding to places of work. This address information was then linked to the Workplaces layer of the National Population Database, using a combination of best-match organisation name matching, and direct postcode matching. The NPD Workplaces layer includes an industry code for each entry, which can be used to derive the broader industry which the workplace belongs.

*S.1.1.6. Commuter age, sex, and ethnicity*
From the 2011 Census, we had data on the age, sex and ethnicity of commuters by workplace zone. We grouped the ages into 4 age groups: 18-29; 30-44; 44-59; 60-64. Ages outside of this range were removed since this is the typical workforce age. Sex was grouped into Male and Female. Ethnicity was grouped into White, Asian, Black/African/Caribbean, and Mixed/Multiple/Other.

For each workzone, we calculated the number of workers in each of these demographic groups. We then calculated the expected proportion of workers in each industry using the NPD workplace layer. We calculated the expected number of workers in each demographic group in each industry by multiplying the number of workers in that demographic group by the proportion of workers in that industry. The data were then aggregated up to MSOA level by summing the number of workers in each demographic group and industry across all workzones. From this aggregated data, the proportion of workers within each demographic group in that industry and MSOA was calculated by dividing by the total number of workers in that industry and MSOA. This was then converted to a percentage.

*S.1.1.7. Commuter method of travel to work*
From the 2011 Census, we had data on the age of commuters by residential MSOA. These data were grouped by mode of travel: Train; Taxi (or vehicle passenger); Single occupancy; Bus, metro or tram; other.

Since these data are at residential MSOA rather than workplace zone, we first needed to map these to workplace zone level, before following the procedure described above for age, sex, and ethnicity. To map these data to workplace zone, we used the travel to work data from the 2011 census. From this, we could calculate the expected number of commuters leaving the focal MSOA to work in the target workplace zone. Multiplying this by the proportion of commuters using each method of transport gives the expected number of commuters using each method of transport who live in the residential MSOA and work in the workzone. Repeating this for all residential MSOAs that commute into the target workzone provides the total expected number of commuters into the target workzone for each method of transport. We then followed the procedure above to aggregate by industry and MSOA.

*S.1.1.8. Commuter IMD*
Using the Department for Levelling Up, Housing and Communities indices of multiple deprivation (IMD) classifications [10], we had data on the age of commuters by residential MSOA. Using this, we calculated the distribution across IMD deciles in each residential MSOA. We then followed the mapping procedure described above for the method of transport data to calculate the expected IMD distribution of commuters into each workplace MSOA. From this, we calculated the mean IMD of commuters into each workplace MSOA.

*S.1.1.9. Workplace mobility class*
The workplace mobility dataset classifies each LSOA into one of eight levels: with level 1 being fully metropolitan, and 8 fully rural. We regroup these into 4 levels: Metropolitan (L1 and L2),

Exurban (L3 and L4), Suburban (L5 and L6), and Rural (L7 and L8). To obtain an MSOA level metric, we defined each MSOA by the most common mobility class amongst LSOAs within that MSOA.

*S.1.1.10. Workplace broad industry sector*
From the NPD workplace layer, we had data on each workplace, providing the name, address, industry sector, and number of employees. For each workzone, we grouped the data by industry sector and counted the number of workplaces and number of employees in that sector within that workzone.

*S.1.1.11. Workplace proportion of employees on permanent contracts*
This data [19] provides the proportion of employees on permanent contracts (permanence) for each two-digit SIC division code (1-99). These SIC groupings are nested within the industry sector definitions used in this study. For each MSOA and industry, we calculated the number of employees within each SIC grouping. We then calculated the average permanence across all SIC groupings within that industry in that MSOA. This was done by multiplying the permanence by the number of employees in each SIC grouping, summing across all SIC groupings within that industry, and then dividing by the number of employees in that industry in that MSOA.

*S.1.1.12. Workplace relative measure of physical proximity*
This data [18] provides a relative measure of physical proximity between employees in the workplace (proximity) for each SIC division code (1-99). This data is derived from the Annual Population Survey, and uses the following proximity scores: 0 - I do not work near other people (beyond 100 ft.); 25 - I work with others but not closely (for example, private office); 50 - Slightly close (for example, shared office); 75 - Moderately close (at arm's length); 100 - Very close (near touching). The data takes the average score for each 2-digit SIC division code. These SIC groupings are nested within the industry sector definitions used in this study. For each MSOA and industry, we calculate the number of employees within each SIC grouping. We then calculate the average proximity across all SIC groupings within that industry in that MSOA. This is done by multiplying the proximity by the number of employees in each SIC grouping, summing across all SIC groupings within that industry, and then dividing by the number of employees in that industry in that MSOA.

*S.1.1.13. Residential characteristics*
In addition to the commuter characteristics, we consider some characteristics of the residential population (individuals who live near to the workplace) as potential adjustment variables. These are: resident IMD quintile, the most common IMD quintile among residents in the workplace MSOA; resident COVID-19 case rate, the proportion of residents in the workplace MSOA testing positive for COVID-19; resident vaccination dose 1, the proportion of residents in the workplace MSOA with 1 or more vaccination dose; resident vaccination dose 2, the proportion of residents in the workplace MSOA with 2 or more vaccination doses.

S.1.2. Identifying adjustment sets for statistical analysis

Figure S1 illustrates the Directed Acyclic Graph (DAG) used to represent predefined hypothetical causal relationships between HSE key priorities relating to industry, employee, and local residential characteristics and clusters of COVID infections. This DAG assumes we are interested in the acute effects of each priority, defined here as their relationship with the following week. The HSE priorities defined prior to the analysis took place were:

Workplace Characteristics
- Location
- Size (Number in Employment)
- Sector (Theme 1 Sector)
    - Employment Type (Temporary/Zero Hours)
    - Younger workers (Employment by Age)
    - Physical Proximity

Employee Travel Characteristics (Commuter)
- Method of travel to work

Employee Demographic Characteristics (socio-economic)
- Indices of Multiple Deprivation
- Ethnicity Group
- Vaccination

These priorities were translated on the Figure S1 DAG into three risk factor groups
1. Workplace Characteristics
2. Commuter Characteristics
3. Mode of Transport

We then proposed the following three adjustment sets for each of the HSE priorities, no adjustment, minimal adjustment, and then HSE priority specific full adjustment. In each case we aim to capture the total association between the HSE priority risk factor and outcome, and so aim to not adjust for characteristics later in the causal path.

1. Unadjusted model, i.e. risk factor only with no adjustment for confounding.
2. Minimal adjusted model, i.e. common confounders to all risk factors including the lagged COVID rate variables, MSOA mobility cluster (local to the workplace), and residential characteristics (local to the workplace).
3. Priority risk factor fully adjusted model, i.e. also adjusted for all variables in the two risk factor groups occurring prior in the hypothetical causal chain to the HSE priority. For example, if we examine a variable in the workplace characteristic risk factor group, we would adjust for employee travel characteristics, and employee residential characteristics. Note, variables within a risk factor group will not be adjusted for due to the complexity of the causal relationships present.

**Figure S1:** *Proposed DAG of hypothetical causal relationships between key MSOA industry characteristics, commuter characteristics, mode of transport, and residential characteristics, and clusters of COVID infections. White variables are the common adjustment set, grey variables*

*are unobserved, and blue variables are our key variables of interest.*

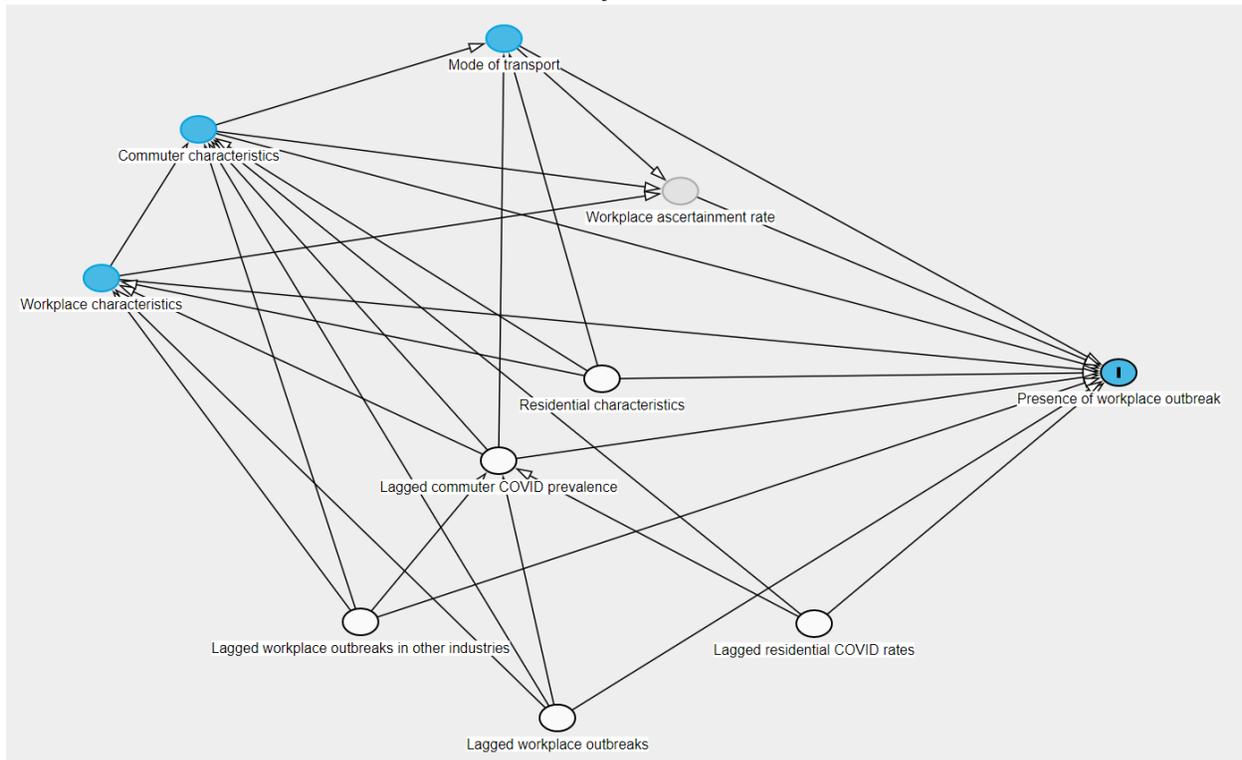

### S.1.3. Bias

The main bias in our data is that different industries will have different testing rates. For example, some industries may have had mandatory testing mandates, requiring all employees to get regularly tested. Therefore, since the clusters are defined in terms of positive tests rather than infections, the probability of clusters being identified are not comparable across industries. To avoid this bias, the primary analysis uses industry-specific models, and we do not compare across industry. In the interpretation of our results, we have converted the effect sizes to instead report the relative increase in outbreak risk from changing the focal variable. Since this is a relative measure, these can be compared across industries.

Another bias arises from the changes in SARS-CoV-2 variants across the study period. At the start of the study period (June 2021), the Delta variant was dominant. However, in Autumn 2021 the Delta Plus variant started to spread [26], before in December 2021 the Omicron variant became dominant. Some of the variables may have differing responses to different variants. To account for this, we considered models over the whole study period and variant specific models. In this study, this bias is expected to have the largest influence on the estimated effect of vaccination. This is because vaccine escape is known to be higher with Omicron than Delta. Since cumulative vaccination is correlated with time, and the number of COVID-19 cases has generally increased across the study window, this is likely to bias the vaccination effect.

Changes to testing policy would affect the analysis conducted. Over the study period, testing policy was reasonably consistent temporally. However, we do not account for spatial variation in testing policy. If some areas have fewer tests at random, this will be adjusted for in the Markov Random Field used to smooth the MSOA level testing data. However, if there is a spatial bias in the testing policy (for example a spatial bias in test availability), this will not be captured by our model. Therefore, both the commuter COVID-19 case rates and the number of outbreak clusters could have some spatial bias, which could be amplified by work from policies [27].

The final major bias is in the permanence and proximity variables. These rely on having sufficient variation within each industry sector in order to be used. However, for some industries, there is insufficient variation, with just a single value for these variables for all MSOAs, or highly correlated permanence and proximity. For such scenarios, these variables do not change enough to accurately measure their influence on workplace outbreaks. To account for this, we only consider these variables for the following industry sectors: Services; Utilities; Transport, distribution and warehousing; Manufacturing; and Construction.

## S.2. Supplementary tables

*Table S1:* Unadjusted model, percentage change in risk of workplace outbreak cluster rate by industry.

| | Percentage change in risk by industry (unadjusted model) | | | | | | | | | |
|---|---|---|---|---|---|---|---|---|---|---|
| Variable | Services | Utilities | Education | Transport, distribution and warehousing | Mining and Quarrying | Manufacturing | Public service activities | Construction | Human health and social work | Waste management and remediation | Agriculture, forestry and fishing |
| **Workplace - proximity** | | | | | | | | | | | |
| Physical proximity in the workplace | 3.5 (3.3,3.8) | 26 (22,30) | NA | 18 (17,19) | NA | 6.6 (6.1,7.1) | NA | -9.7 (-11,-8.1) | NA | NA | NA |
| **Workplace - permanence** | | | | | | | | | | | |
| Proportion of workers on permanent contracts | 0.22 (-0.37,0.81) | -27 (-31,-23) | NA | -43 (-44,-41) | NA | -20 (-21,-20) | NA | -43 (-56,-27) | NA | NA | NA |
| **Workplace - mobility class** | | | | | | | | | | | |
| Mobility class - Exurban | -30 (-31,-28) | -9.8 (-21,2.9) | -13 (-14,-11) | -16 (-19,-12) | 93 (38,1.7e+02) | -13 (-15,-10) | -38 (-41,-35) | -31 (-36,-25) | -12 (-14,-9.6) | 5.1 (-9,21) | -26 (-37,-12) |
| Mobility class - Metropolitan | -18 (-19,-16) | -39 (-45,-32) | -10 (-12,-8.4) | -15 (-17,-12) | -74 (-85,-52) | -15 (-18,-12) | 13 (8.8,17) | 0.079 (-6,6.6) | -11 (-13,-9.4) | 52 (33,74) | 7e+02 (1.5e+02,4.7e+03) |
| Mobility class - Rural | -32 (-34,-30) | -31 (-44,-15) | -53 (-55,-52) | -17 (-22,-13) | 15 (-21,68) | -17 (-21,-14) | 16 (8,25) | -37 (-44,-29) | -6.2 (-10,-1.8) | 52 (26,84) | -71 (-76,-65) |
| Mobility class - Suburban | baseline | baseline | baseline | baseline | baseline | baseline | baseline | baseline | baseline | baseline | baseline |
| **Commuter - vaccination** | | | | | | | | | | | |
| Proportion of commuters with 2 or more vaccination doses - Delta | -1.6 (-1.7,-1.5) | -2.6 (-3.5,-1.7) | 3.1 (2.9,3.2) | -1.1 (-1.3,-0.83) | 3.9 (0.84,7) | -1.1 (-1.3,-0.84) | -0.3 (-0.66,0.059) | -2.9 (-3.4,-2.4) | 0.56 (0.36,0.75) | -2.6 (-3.8,-1.4) | -0.53 (-2.1,1.1) |
| Proportion of commuters with 2 or more vaccination doses - Omicron | -2.7 (-2.9,-2.5) | -4.7 (-6.2,-3.1) | 0.5 (0.27,0.73) | -3.1 (-3.5,-2.7) | 2.3 (-3.7,8.6) | -1.7 (-2.1,-1.3) | -4.2 (-4.8,-3.6) | -4.8 (-5.8,-3.9) | 0.59 (0.34,0.83) | -0.11 (-1.9,1.7) | -17 (-19,-14) |
| **Commuter - age** | | | | | | | | | | | |
| Proportion of commuters aged 18-29 | 3 (2.7,3.2) | 3.6 (2.4,4.7) | -0.66 (-0.96,-0.37) | 1.2 (0.73,1.7) | -0.78 (-4.6,3.2) | -1.8 (-2.3,-1.3) | -0.81 (-1.4,-0.26) | -0.58 (-1.7,0.57) | -4.5 (-4.9,-4.2) | 5.4 (3.3,7.6) | 3 (0.31,5.7) |
| Proportion of commuters aged 30-44 | baseline | baseline | baseline | baseline | baseline | baseline | baseline | baseline | baseline | baseline | baseline |
| Proportion of commuters aged 45-59 | 6.8 (6.4,7.1) | 9 (7.7,10) | 2.7 (2.4,3) | 6.5 (5.9,7) | 13 (9.2,18) | 5.5 (5,6) | -1.1 (-1.6,-0.51) | 2.7 (1.5,3.8) | 3.1 (2.7,3.5) | 7.9 (5.8,10) | -3.4 (-6.5,-0.13) |
| Proportion of commuters aged 60-64 | -11 (-11,-10) | -9.3 (-12,-6.9) | -6.9 (-7.3,-6.4) | -14 (-15,-13) | -16 (-21,-12) | -20 (-21,-20) | -8.1 (-9.2,-7) | -13 (-15,-11) | -10 (-11,-9.5) | -3.1 (-6.2,0.21) | -20 (-25,-16) |
| **Commuter - ethnicity** | | | | | | | | | | | |
| Proportion of commuters with an asian ethnicity | 0.044 (-0.073,0.16) | 4.4 (3.2,5.5) | -0.22 (-0.35,-0.094) | -1.1 (-1.4,-0.92) | 4.6 (0.64,8.8) | -0.66 (-0.88,-0.44) | -0.17 (-0.48,0.14) | 0.41 (-0.13,0.96) | 0.59 (0.44,0.74) | 0.029 (-1.1,1.2) | 3.7 (1.6,5.8) |
| Proportion of commuters with a black/african/caribbean ethnicity | 1.7 (1.4,2) | 6.9 (4,9.9) | -1.4 (-1.7,-1.1) | 3.4 (2,8.4) | 14 (-13,48) | 1.4 (0.47,2.3) | 1 (0.49,1.5) | 1.9 (0.6,3.3) | 3.4 (3.1,3.7) | 7.9 (5,11) | -0.94 (-9.2,8.1) |
| Proportion of commuters with a mixed/multiple/other ethnicity | -9.4 (-10,-8.7) | -35 (-38,-31) | 0.62 (-0.11,1.4) | -7.3 (-8.7,-5.8) | -52 (-65,-33) | -11 (-12,-8.8) | 7.2 (5,4.9) | -5 (-7.9,-1.9) | -13 (-14,-12) | -11 (-14,-7.4) | 84 (58,1.1e+02) |
| Proportion of commuters with a white ethnicity | baseline | baseline | baseline | baseline | baseline | baseline | baseline | baseline | baseline | baseline | baseline |
| **Commuter - sex** | | | | | | | | | | | |
| Proportion of commuters who identify as Female | -0.44 (-0.53,-0.35) | -0.66 (-1,-0.27) | 3 (2.9,3.1) | -1.4 (-1.6,-1.3) | -5.1 (-6,-4.1) | -3.5 (-3.6,-3.3) | -1.5 (-1.7,-1.4) | -2.8 (-3.1,-2.4) | 3.1 (3,3.3) | 0.92 (0.36,1.5) | 0.14 (-1.3,1.6) |
| Proportion of commuters who identify as Male | baseline | baseline | baseline | baseline | baseline | baseline | baseline | baseline | baseline | baseline | baseline |
| **Commuter - IMD** | | | | | | | | | | | |
| Commuter IMD quintile | -14 (-15,-13) | -26 (-31,-21) | 3.5 (2.4,4.5) | -8.7 (-10,-7) | -42 (-53,-27) | -15 (-17,-14) | -4.5 (-6.8,-2.1) | -23 (-26,-20) | -7.8 (-8.9,-6.6) | -3.7 (-11,3.7) | -10 (-19,0.15) |
| **Method of travel to work** | | | | | | | | | | | |
| Proportion of commuters using bus/metro/tram | 0.43 (0.35,0.52) | -0.6 (-1.1,-0.059) | -0.71 (-0.82,-0.6) | 0.77 (0.57,0.97) | -5.2 (-9.1,-1.1) | 0.81 (0.55,1.1) | 0.17 (-0.048,0.39) | -0.13 (-0.51,0.26) | 0.84 (0.71,0.97) | 3.2 (2,4.5) | 1.4 (-1.1,3.9) |
| Proportion of commuters using taxi/vehicle passenger | 2 (0.029,4.1) | -13 (-24,0.89) | -18 (-20,-17) | -11 (-14,-7.2) | -51 (-63,-35) | -35 (-38,-32) | 15 (11,20) | -30 (-37,-22) | -16 (-18,-13) | 12 (-3.7,31) | -29 (-40,-15) |
| Proportion of commuters using other transport | baseline | baseline | baseline | baseline | baseline | baseline | baseline | baseline | baseline | baseline | baseline |
| Proportion of commuters using train | 8.8 (8.4,9.2) | -0.49 (-2.2,1.3) | -3.5 (-3.9,-3) | 5.7 (5.1,6.3) | -11 (-16,-6) | 7.1 (6.6,7.7) | -6.1 (-6.9,-5.2) | 7.2 (5.6,8.9) | 0.92 (0.4,1.4) | -2.6 (-4.8,-0.38) | 14 (12,17) |
| Proportion of commuters using single occupancy | -0.91 (-1.1,-0.73) | -5.9 (-6.6,-5.1) | -1.6 (-1.8,-1.3) | -1.3 (-1.6,-0.98) | -2.9 (-8.1,2.5) | -6.9 (-7.5,-6.3) | -0.95 (-1.2,-0.68) | 1.8 (1.3,2.4) | -4.1 (-4.4,-3.8) | -3.3 (-5.7,-0.86) | 9.2 (5.7,13) |

*Table S2:* Minimal adjusted model, percentage change in risk of workplace outbreak cluster rate by industry. NA values for workplace characteristics since there is no minimal adjusted model for these

*variables.*

| | Percentage change in risk by industry (partially adjusted model) | | | | | | | | | | |
|---|---|---|---|---|---|---|---|---|---|---|---|
| Variable | Services | Utilities | Education | Transport, distribution and warehousing | Mining and Quarrying | Manufacturing | Public service activities | Construction | Human health and social work | Waste management and remediation | Agriculture, forestry and fishing |
| **Workplace - proximity** | | | | | | | | | | | |
| *Physical proximity in the workplace* | NA | NA | NA | NA | NA | NA | NA | NA | NA | NA | NA |
| **Workplace - permanence** | | | | | | | | | | | |
| *Proportion of workers on permanent contracts* | NA | NA | NA | NA | NA | NA | NA | NA | NA | NA | NA |
| **Workplace - mobility class** | | | | | | | | | | | |
| *Mobility class - Exurban* | NA | NA | NA | NA | NA | NA | NA | NA | NA | NA | NA |
| *Mobility class - Metropolitan* | NA | NA | NA | NA | NA | NA | NA | NA | NA | NA | NA |
| *Mobility class - Rural* | NA | NA | NA | NA | NA | NA | NA | NA | NA | NA | NA |
| *Mobility class - Suburban* | NA | NA | NA | NA | NA | NA | NA | NA | NA | NA | NA |
| **Commuter - vaccination** | | | | | | | | | | | |
| *Proportion of commuters with 2 or more vaccination doses - Delta* | -3.8 (-4.1,-3.5) | -2.8 (-4.9,-0.76) | 1.7 (1.2,2.1) | -3.6 (-4.2,-3) | 7.5 (-3.6,20) | -2.3 (-2.9,-1.8) | -3.3 (-4.1,-2.5) | -1.1 (-2.4,0.22) | -2 (-2.5,-1.6) | -2.3 (-5.4,0.86) | -6 (-11,-0.67) |
| *Proportion of commuters with 2 or more vaccination doses - Omicron* | -3.3 (-3.6,-3.1) | -7.1 (-9.1,-5.1) | -0.2 (-0.55,0.14) | -4.4 (-4.9,-3.9) | 8 (-0.36,17) | -2.3 (-2.8,-1.7) | -6.1 (-6.8,-5.4) | -0.44 (-1.8,0.97) | -2.5 (-2.8,-2.2) | 0.62 (-1.9,3.2) | -11 (-14,-7.1) |
| **Commuter - age** | | | | | | | | | | | |
| *Proportion of commuters aged 18-29* | 1 (0.83,1.2) | 5.5 (4.3,6.8) | -0.2 (-0.48,0.072) | 3.2 (2.8,3.7) | 1.7 (-2.7,6.2) | 0.19 (-0.29,0.66) | 0.76 (0.21,1.3) | 1 (-0.16,2.2) | -2.5 (-2.8,-2.1) | 4.2 (2,6.5) | 1.2 (-1.5,3.9) |
| *Proportion of commuters aged 30-44* | baseline | baseline | baseline | baseline | baseline | baseline | baseline | baseline | baseline | baseline | baseline |
| *Proportion of commuters aged 45-59* | 5.3 (5,5.6) | 9.1 (7.8,10) | 2.1 (1.8,2.4) | 6.9 (6.3,7.4) | 13 (8.3,18) | 4.2 (3.7,4.6) | -0.47 (-1.1,0.13) | 4.5 (3.4,5.6) | 2.2 (1.8,2.5) | 5.2 (3.1,7.3) | -2.5 (-5.7,0.71) |
| *Proportion of commuters aged 60-64* | -4.8 (-5.3,-4.2) | 1.4 (-1.3,4.2) | -4.3 (-4.8,-3.8) | -6.3 (-7.2,-5.4) | -11 (-16,-6) | -12 (-13,-11) | -3.6 (-4.8,-2.4) | -5.2 (-7.2,-3.1) | -4.1 (-4.7,-3.5) | -0.45 (-3,8,3) | -15 (-19,-10) |
| **Commuter - ethnicity** | | | | | | | | | | | |
| *Proportion of commuters with an asian ethnicity* | 0.19 (0.075,0.3) | 2 (0.84,3.1) | -0.056 (-0.18,0.069) | -1.1 (-1.3,-0.89) | 3.5 (-1.1,8.2) | -0.51 (-0.72,-0.31) | 0.29 (-0.033,0.61) | 1.3 (0.71,1.9) | 0.59 (0.44,0.74) | 1.3 (0.089,2.5) | 3.1 (0.85,5.4) |
| *Proportion of commuters with a black/african/caribbean ethnicity* | 1.8 (1.6,2.1) | 6.6 (4.6,8.7) | 0.38 (0.087,0.68) | 3.8 (3.2,4.5) | 7.8 (-19,44) | 2.1 (1.2,3) | 0.22 (-0.32,0.75) | 0.98 (-0.32,2.3) | 3 (2.7,3.3) | 5.5 (2.5,8.6) | -4.3 (-12,4.4) |
| *Proportion of commuters with a mixed/multiple/other ethnicity* | -9.7 (-10,-9.1) | -33 (-36,-29) | -2.3 (-3,-1.6) | -7.9 (-9.3,-6.4) | -45 (-63,-18) | -8.5 (-10,-6.8) | 6.5 (4.6,8.4) | -13 (-15,-9.7) | -11 (-12,-9.8) | -9.4 (-16,-2.9) | 78 (52,1.1e+02) |
| *Proportion of commuters with a white ethnicity* | baseline | baseline | baseline | baseline | baseline | baseline | baseline | baseline | baseline | baseline | baseline |
| **Commuter - sex** | | | | | | | | | | | |
| *Proportion of commuters who identify as Female* | -0.63 (-0.71,-0.54) | -0.69 (-1.1,-0.3) | 1.8 (1.7,1.9) | 0.5 (0.34,0.65) | -3.1 (-4.2,-2.1) | -1 (-1.2,-0.87) | -1.1 (-1.3,-0.99) | -1.4 (-1.8,-1.1) | 1.2 (1.1,1.3) | 1.2 (0.6,1.7) | 1.8 (0.39,3.2) |
| *Proportion of commuters who identify as Male* | baseline | baseline | baseline | baseline | baseline | baseline | baseline | baseline | baseline | baseline | baseline |
| **Commuter - IMD** | | | | | | | | | | | |
| *Commuter IMD quintile* | -19 (-20,-18) | -31 (-37,-26) | -6.2 (-7.4,-5) | -15 (-17,-13) | -15 (-33,9.8) | -14 (-16,-13) | -14 (-16,-11) | -21 (-25,-17) | -19 (-20,-17) | -11 (-19,-1.2) | -7.3 (-19,6) |
| **Method of travel to work** | | | | | | | | | | | |
| *Proportion of commuters using bus/metro/tram* | 0.26 (0.17,0.34) | -0.81 (-1.4,-0.2) | -0.11 (-0.23,0.0039) | 1.2 (1,1.4) | -4 (-8,8.1) | 0.71 (0.44,0.99) | 0.39 (0.14,0.65) | -0.54 (-0.94,-0.13) | 1.2 (1.1,1.3) | 3.2 (1.9,4.5) | 4.5 (1.9,7.2) |
| *Proportion of commuters using taxi/vehicle passenger* | 28 (26,30) | 17 (2.2,34) | -4.5 (-6.2,-2.8) | 10 (7.4,13) | -41 (-56,-22) | -5.5 (-9.2,-1.7) | 14 (9.6,19) | -1.5 (-11,9.1) | 9.5 (7.2,12) | 15 (-1.7,35) | -27 (-39,-13) |
| *Proportion of commuters using other transport* | baseline | baseline | baseline | baseline | baseline | baseline | baseline | baseline | baseline | baseline | baseline |
| *Proportion of commuters using train* | 5.8 (5.4,6.2) | -2 (-3.7,-0.24) | -2.2 (-2.6,-1.8) | 3.6 (3.1,4.2) | -9.3 (-15,-3.3) | 3.6 (3.2,4.1) | -4.6 (-5.5,-3.6) | 4.4 (2.8,6) | 2.9 (2.4,3.4) | -2 (-4.3,0.27) | 7.9 (5.8,10) |
| *Proportion of commuters using single occupancy* | -1.4 (-1.6,-1.3) | -6.1 (-6.8,-5.5) | -3 (-3.2,-2.8) | -2.6 (-2.9,-2.3) | -3.7 (-10,3.1) | -4.5 (-5,-3.9) | -1.6 (-1.9,-1.4) | 0.042 (-0.44,0.52) | -3.7 (-3.9,-3.5) | -3.1 (-5.5,-0.71) | 3.1 (-0.2,6.6) |